\documentclass[reqno]{amsart}
\usepackage{graphicx,amsmath}
\usepackage{natbib}
\usepackage{color}

\newcount\n  \newdimen\w

\def\Repeat#1#2{\n=#1\relax\loop\ifnum       
  \n>0\relax #2\advance\n by-1\repeat}

\long\def\OMIT#1{\relax }  

\def\re#1{(\ref{#1})}   
  
\def\eqn#1#2{ \begin{align} \label{#1}         #2 \end{align}}

\def\nl#1{          \\ \label{#1}        }  

\def\delim#1#2#3{\csname\ifcase#1 relax\or   
   big\or Big\or bigg\or Bigg\fi\endcsname   
  {\ifcase#2\or\Delim#3\or\deliM#3\fi}}      
\def\Delim#1{\ifcase#1\relax\or(\or[\or\{\or<\or\langle\or|\or\|\or---{ }\fi}
\def\deliM#1{\ifcase#1\relax\or)\or]\or\}\or>\or\rangle\or|\or\|\or{ }---\fi}

\let\f\frac                     
        
\def\largerfrac#1#2#3{      
  \whichtypesize\n=\currenttypesize\advance\n by #1 \mathchoice
  {\setbox0\hbox{$\displaystyle-$} \w=.5\ht0\advance\w by-.5\dp0\setbox0
    \hbox{\typesize\n $\displaystyle-$} \advance\w by -.5\ht0\advance\w
    by .5\dp0\raise\w \hbox{\typesize\n$\displaystyle{\frac{#2}{#3}}$}}
  {\setbox0\hbox{$-$} \w=.5\ht0 \advance\w by -.5\dp0 \setbox0\hbox
    {\typesize\n $-$} \advance\w by-.5\ht0\advance\w by
    .5\dp0\raise\w\hbox{\typesize\n$\frac{#2}{#3}$}}
  {\setbox0\hbox{$\scriptstyle-$} \w=.5\ht0 \advance\w by-.5\dp0\setbox0
    \hbox{\typesize\n $\scriptstyle-$} \advance\w by -.5\ht0 \advance\w
    by .5\dp0 \raise\w\hbox{\typesize\n$\scriptstyle{\frac{#2}{#3}}$}}
  {\setbox0\hbox{$\scriptscriptstyle-$} \w=.5\ht0
    \advance\w by -.5\dp0 \setbox0\hbox{\typesize\n
    $\scriptscriptstyle-$} \advance\w by -.5\ht0 \advance\w by .5\dp0
    \raise\w\hbox{\typesize\n$\scriptscriptstyle{\frac{#2}{#3}}$}}  }

\begin{document}

\title{Thermodynamic aspects of rock friction}
\author{N. Mitsui$^{1}$ and P. V\'an$^{1,2}$ }
\address{$^1$Dept. of Theoretical Physics, Wigner RCP, RMKI, \\  H-1525 Budapest, P.O.Box 49, Hungary; 
and  {$^2$Dept. of Energy Engineering, Budapest Univ. of Technology and Economics},\\
  H-1111, Budapest, Bertalan Lajos u. 4-6,  Hungary}

\date{\today}

\begin{abstract}
Rate- and state-dependent friction law for velocity-step tests is analyzed from a thermodynamic point of view. A simple macroscopic non-equilibrium thermodynamic model with a single internal variable reproduces instantaneous jump and relaxation. Velocity weakening appears as a consequence of a plasticity related  nonlinear coefficient. Permanent part of displacement corresponds to plastic strain, and relaxation effects are analogous to creep in thermodynamic rheology.
\end{abstract}

\maketitle

\section[intro]{Introduction}

Frictional force appears between two contacting objects and it influences their relative motion. In geophysics, relative motion at plate boundary is considered as frictional sliding. The frictional interaction of continental plates leads to earthquakes. Large earthquakes generate propagating coseismic deformations as seismic waves and they can cause disasters, therefore it would be useful to know when, where, and how large earthquakes will occur in order to mitigate the damages by appropriate measures. This is a difficult problem.

The difficulty partially comes from the incomplete understanding of earthquake mechanism. Earthquakes are caused by fracture and deformation in the earth'’s crust. Therefore the deformation before, and the fracture during an earthquake have to be modelled. One may think that the difficulty of earthquake mechanism comes from the instability of the fracture. However, the incompleteness of the deformation theory of the medium with external force is also essential. For example, crystallized rock material of the crust is fragmented, and grains with various size (e.g., sand, mud) may influence the mechanical properties, too. Consequently, the theory should explain the deformation by the applied external stress based on a constitutive law (with temperature if possible). However, present theory is based mainly on elasticity (e.g., \citealt{AkiRic02b}) and the constitutive laws are empirical.

The relation between friction and earthquake can be understood with the example of metals. In tribology, one cannot avoid erosion of frictional surfaces even if they are polished and smooth. The wear off the surface, the erosion, is related to the accumulation and the release of stress, which leads to waves and vibrations. Thus, the mechanism of earthquakes is connected to the instability of the friction laws.

Properties of rock friction have been investigated mainly by laboratory experiments, and empirical equations have been proposed as rate- and state-dependent friction law (\citealt{Die79a}; \citealt{Rui83a}). There are various models to explain the mechanism (\citealt{BauCar06a}; \citealt{PutEta11a}), developing former micro- and mesoscopic ideas as the real contact area theory (\citealt{BowTab50b}) and thermal activation theory (\citealt{Eyr36a}). 

Thermal activation theory is originally proposed for chemical reactions in 
\cite{Eyr35a}. { It was first adopted to friction in \citealt{HesEta94a} for 
paper, and in \citealt{BreEst94a} for metals. First was applied for rock 
friction in \citealt{Nak01a}, and later analysed in \citealt{BauCar06a} and 
\citealt{PutEta11a} through the absolute rate theory about creep of crystals 
(\citealt{Pol85b}).} Rock friction is considerably related to the erosion of 
surfaces and the degradation and wear of rock fragments between the sliding 
samples. Therefore it is not evident that microchemistry is the best approach 
to grasp the universal aspects of the background mechanism. Experimental 
results of granular friction show similar features to those of bare rocks, this 
way supporting our opinion (\citealt{Maro98a}; \citealt{KawEta12a}). 

In what follows, we focus on the universal aspects of friction, avoiding micro- and mesoscopic considerations. These aspects are best understood when only macroscopic concepts are applied, those which are expected to be valid independently of microstructures and micromechanisms. From this point of view, the second law of thermodynamics is the straightforward theoretical background for rate- and state-dependent friction. We consider the dynamic, non-equilibrium, interpretation and extension of thermodynamics of homogeneous bodies, where the conditions of stability of equilibrium are connected to the second law (\citealt{Mat05b}). Stability is a key aspect in rate- and state-dependent friction, too.

In this paper, we treat the dependence of dynamic friction on the shear loading velocity (\citealt{Die79a}; \citealt{Rui83a}) in the framework of non-equilibrium thermodynamics of homogeneous (discrete) bodies with internal variables (\citealt{MauMus94a1,MauMus94a2}). First, we summarize the qualitative properties of rock friction experiments. Second, we mention the key points of thermodynamical modeling and derive the constitutive differential equations in Section 3. After that we discuss the results and compare the model with experimental expectations in Section 4.

\section{Summary of rock experiments and empirical constitutive laws}

The rock experiments of sliding friction are described by the so-called rate- and state-dependent friction law. The equations of this law unify the results obtained from two types of rock experiments; the first one is the time dependence of static coefficient of friction (\citealt{Die72a}) and the second one is slip velocity dependence of the dynamic coefficient of friction (\citealt{Die78a}).

The properties of dynamic friction are the following (see the details in \citealt{Maro98a}): 
\begin{enumerate}
 \item frictional coefficient in stable sliding conditions with a constant load-point velocity depends on the logarithm of the load-point velocity;
 \item the magnitude of the instantaneous jump of the frictional coefficient depends on the change of the load-point velocity;
 \item the following relaxation of the frictional coefficient to new value in stable sliding is also dependent on the instantaneous change of the load-point velocity; 
 \item oscillation occurs in some cases (e.g., large load-point velocity, polished surfaces, thin sand interface layer between the samples) (e.g., \citealt{MarEta90a}).
\end{enumerate}
  
These properties can be reproduced by using two classical equations except the oscillation. The first one is the constitutive law \re{Die_flaw}, expressing the relation between frictional coefficient $\mu$ and slip velocity $V$ with an additional variable, called state variable $\theta$. The second one is the evolution law \re{Die_elaw} expressing the time evolution of state variable depending on the slip velocity. The followings are by \cite{Die79a}.
\begin{eqnarray}
 \mu = \frac{\tau}{\sigma} &=& 
    \mu_* + a\ln\left( \frac{V}{V_*} \right)\ + b\ln\left( \frac{V_{*}\theta}{D_c} \right), \label{Die_flaw}\\
 \frac{d\theta}{dt} &=& 
    1-\frac{V\theta}{D_c},
\label{Die_elaw}\end{eqnarray}
where $\tau$ is the shear stress, $\sigma$ is the normal stress, $\mu_{*}$ is the  constant frictional coefficient for steady-state slip at reference slip velocity $V_{*}$, $a$ and $b$ are the material parameters, $D_{c}$ is the critical slip distance, and $t$ is time. Later on an other evolution law was proposed by \cite{Rui83a}.\eqn{Rui_elaw}{
 \frac{d\theta}{dt} = 
    -\frac{V\theta}{D_c} \ln\left(\frac{V\theta}{D_c} \right).
}

Experimental data of static friction is better reproduced by the equations of \cite{Die79a} (eqs. \re{Die_flaw}  and \re{Die_elaw}), and the dynamic one is by those of \cite{Rui83a} (eqs. \re{Die_flaw} and \re{Rui_elaw}) (see the comparison with experimental data in  \citealt{Maro98a}). Thus another versions have been proposed (e.g., \citealt{PerEta95a}; \citealt{KatTul01a}) in order to reproduce the experimental data better. However, none of them are completely satisfactory.

There are two particular problematic aspects in these laws; the first one is, that the meaning of the state variable $\theta$ is not clear, and the second one is that slip velocity at the frictional surface is assumed to be equal to the load-point velocity of the contacted rock samples. These observations lead to the following improvements in a recent version (\citealt{NagEta12a}):
\eqn{Nag_flaw}{
 \tau &= \Theta+ a\sigma\ln\left( \frac{V}{V_*} \right), \nl{Nag_elaw}
 \frac{d\Theta}{dt} &= \frac{b\sigma}{D_c/V_*} \exp\left(- 
 \frac{\Theta-\Theta_*}{b\sigma} \right)\ 
      -\frac{b\sigma}{D_c}V- c\frac{d\tau}{dt} .
}
where $\Theta_*$ is the shear strength for steady-state slip at $V_*$, and $c$ is a positive constant. In these equations, state variable $\theta$ is replaced by shear strength $\Theta$ according to \cite{Nak01a}, and slip velocity $V$ is different from load-point velocity $V_l=V+\frac{1}{k}\frac{d\tau}{dt}$, estimated by another way with shear stress $\tau$ and system stiffness $k$. The third term in the right hand side of \re{Nag_elaw} is a stress weakening term, a proposition based on the estimation of shear strength by the acoustic wave transmissivity experiment of \cite{NagEta12a}. These equations reproduce also oscillation of the shear stress and fit the data better than the previous versions. 

{ The above mentioned equations reproduce several aspects of the experimental 
results, but their origin is empirical, and their form is not explained. The 
application of thermal activation theory was an important step toward an 
explanation.  \cite{Nak01a} partially verified the theory in rock friction 
experiments. However, the direct influence of thermal activation for mechanical 
phenomena is not evident. This is also indicated by the observation, that in 
experiments in \cite{NagEta12a} the rock sample was prepared to level out of 
the contacted surface by sliding (Nagata, personal communication).}

\section{Thermodynamics of the frictional layer}

Stability and dissipation are two characteristic concepts that require some clarification before considering the specific properties of friction.

\subsection{Thermodynamic approach}

Thermodynamic {\em processes} can be classified as equilibrium, steady-state, and non-equilibrium. Related concepts are equilibrium and non-equilibrium  thermodynamic {\em states}. Equilibrium thermodynamic bodies are described only by state variables that may be nonzero in case of equilibrium processes. On the other hand, in state space of non-equilibrium bodies there are variables that have zero values in case of equilibrium processes. These are called internal variables in the following in the sense of \cite{Ver97b}. Our thermodynamic model of friction assumes homogeneous thermodynamic bodies, where continuum and other structural aspects of the phenomena are given by additional variables in the homogeneous modelling framework (\citealt{Mat05b}; \citealt{CimEta14a}) of rate- and state-dependent friction. The aforementioned properties of dynamic friction are regarded as follows; 1) stable sliding is steady-state, and 2)-4) shear stress reaction in velocity-step test and oscillation is non-
equilibrium process. 
The state variables $\theta$ and $\Theta$ in the rate- and state-dependent friction models resemble an internal variable from a thermodynamic point of view. 

Dissipation can be interpreted by the nonvanishing source term in the entropy balance. It is directly related to the difference between equilibrium and non-equilibrium. Here we have to consider carefully the relation between dissipation and friction. Dissipation is irreversible phenomenon, and permanent displacement corresponds to irreversible phenomenon in rock friction. On the other hand, the recoverable part of displacement does not lead to dissipation, it  is analogous to elastic deformation in solids. Consequently we divide frictional displacement into irreversible and reversible parts.

The entropy of the body, $S$, is the function of the internal energy $U$ and the additional state and internal variables. The partial derivative of entropy by the internal energy is the reciprocal temperature, 1/$T$. The  stability of physical systems is related to the existence of entropy of a body as a thermodynamic potential and  the the requirement that the total entropy is increasing along thermodynamic processes in insulated thermodynamic systems in general (\citealt{Mat05b}).  This is the background of our following simple dynamic friction models.

\subsection{Macroscopic model of rock friction in thermodynamics }

In the following we develop a treatment of sliding friction in the framework of 
non-equilibrium thermodynamics. The particular approach has three independent 
parts. First of all the thermodynamic theory of rheology with internal 
variables motivates the evolution equations of internal variables that are 
compatible with the second law (\citealt{Ver97b}). Secondly, permanent 
displacement is introduced by a rate variable, as it is necessary in case of 
plasticity theories (see e.g., \citealt{FulVan12a}). Finally the friction 
related a part of entropy production is not quadratic, but first order Euler 
homogeneous expression of the thermodynamic forces, analogous to plasticity 
theories (\citealt{Zie81b}; \citealt{Mau99b}). The particular nonlinear 
function is motivated by the non-associative plasticity concept of 
\cite{HouPuz06b} with the help of  nonlinear Onsagerian conductivity 
coefficients (\citealt{Van10a1}).

\begin{figure}
\begin{center}
\includegraphics[scale=0.4]{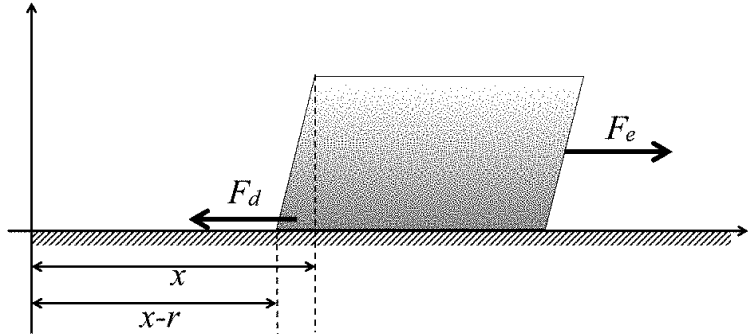}
\end{center}
\caption{ \label{fig_slidingbody}
 Sliding thermomechanical body
}
\end{figure}

Let us consider a body on a horizontal surface with mass $m$. There are two forces that determine the motion of the body: the external force $F_e$, and the damping force $F_d$, due to friction (Fig. \re{fig_slidingbody}). The position of the body is denoted by $x$. The body is not considered completely rigid, however we assume that one particular material point of the body may characterize its instantaneous position. The equation of motion is
\eqn{Newt}{
  m\ddot x = F_e - F_d.
}
Moreover, we assume that the work of the external force changes the energy of the body, $E$. Therefore  
\eqn{Etot}{
  \dot E = F_e \dot x.
}

In this case thermodynamics requires that the damping force contributes only to the internal energy of the body. Our body is an open system energetically, however, we do not calculate directly the energy balance of the environment here. Instead of it, we assume that the external force accelerates the body and also that the body is deformable. In our homogeneous model, the deformation is expressed by an internal variable, the recoverable displacement, $r$. Therefore we distinguish between kinetic and the elastic energies of the body.

The internal energy, $U$, is the difference of the total energy, $E$, the kinetic energy and the elastic energy:
\eqn{Eint}{
U=E-m\frac{(\dot x)^2}{2}-k\frac{r^2}{2},
}
where  $k$ is the parameter of the elasticity, $r$ is the recoverable part of the displacement and we assume a particular kinematic condition: the instantaneous position of the body is the sum of a permanent and a recoverable displacements. A convenient method of their distinction is an additive separation of the displacement rates:
\eqn{plastkin}{
  V= \dot x = \dot r + z,
}
where $V$ is the rate of the position $x$, and $z$ is the rate of the permanent displacement. This kinematical condition is well accepted method to introduce the distinction of plastic and elastic strain rates in plasticity (see e.g., \citealt{FulVan12a}; \citealt{RusRus11b}). 

The dissipation can be calculated by the entropy balance, assuming that the entropy is the function of the internal energy only:
\eqn{Sbal}{
\dot S(U)=\frac{1}{T}\left( F_e \dot x -mV\dot V-kr\dot r  \right)\geq0\,\ \Rightarrow\,T\dot S =F_d V-kr\dot r\geq0.
}
The damping force and also the rate of the recoverable displacement $\dot r$ 
are the constitutive quantities to be determined in accordance with the 
requirement of nonnegative entropy production. { The entropy produced by 
friction is dissipated to the environment as heat. The stationary temperature 
of the sliding bodies depends on the efficiency of the heat transfer between 
the body, the surface and the environment. The above inequality does not refer 
to this process, here the entropy balance is related to the body and not to the 
environment.}

The simplest solution of the inequality \re{Sbal} assumes linear relationships between the thermodynamic fluxes and forces with coefficients $L_1$, $L_2$, $L_{12}$, and $L_{21}$ (\citealt{GroMaz62b}). Therefore
\eqn{lin1}{
 F_d &= L_1 \dot x - L_{12}kr,}
 \eqn{lin2}{ 
 \dot r &= L_{21}\dot x -L_2 kr.
}

The left hand sides show the thermodynamic fluxes and they are related to thermodynamic forces with thermodynamic conductivity coefficients in the right hand sides. The symmetric part of the matrix of the constant thermodynamic coefficients is positive definite, therefore 
\eqn{tineq}{
L_1>0, \qquad L_2>0, \quad \text{and} \quad  L_1 L_2-L^2\geq0, 
}
where $L= (L_{12} +L_{21})/2$. Substituting eq. \re{Newt} to eq. \re{lin1} we obtain
\eqn{lins1}{
  m\ddot x &= F_e -L_1 \dot x + L_{12}kr, \nl{lins2}
  \dot r &= L_{21}\dot x - L_2 kr.
}
We can eliminate the recoverable displacement $r$ and obtain a third order differential equation for $x$:
\eqn{frirheo1}{
\dot F_e + L_2 k F_e= m\dddot x+\left( L_{1}+L_2 km \right)\ddot x +\left( L_1 L_2 -L^2 + a^{2}\right) k \dot x  .
}

Equivalently, displacement $x$ can be eliminated and  the recoverable displacement is obtained as independent variable of the differential equation:
\eqn{frirheo2}{ 
 m\ddot r + \left( L_{1}+L_2 km \right)\dot r = (L-a) F_e  - \left( L_1 L_2 -L^2 +a^{2}\right) kr  
}
where $a = (L_{12} -L_{21})/2$. In these equations the coefficient $L_1$ 
characterizes direct mechanical damping effects. If it is constant, then it can 
be identified as the classical damping which is proportional in magnitude and 
opposite in direction to the velocity. Equation \re{frirheo1} can be rewritten 
introducing the velocity $V=\dot x$ instead of the displacement: {
\eqn{frirheo}{
\tau_1 \dot F_e + F_e= \tau_2 \hat m\ddot V + \hat m \dot V + \eta V,
}
where we have introduced new coefficients 
\eqn{ncoeff}{
\tau_1= \f{1}{L_2 k}, \quad 
\hat m = m+ \f{L_1}{L_2 k}, \quad 
\tau_2 = \frac{m}{L_1+ m k L_2}, \quad 
\eta = L_1 - \f{L_{12}L_{21}}{L_2}.
} 

These material parameters are nonnegative according to the thermodynamic requirements in \re{tineq}.} We can recognise different relaxation modes for the force and the velocity, depending on the mode of the experimental control. When the velocity is controlled, by abruptly changing between two stationary states, then the force is relaxed with relaxation time $\tau_1$ between the corresponding stationary values. When one controls the force, then the velocity is ``creeping'' between stationary values with relaxation time $\tau_2$ and inertia $\hat m$. This is analogous to the standard model of rock rheology, the so called Poynting-Thomson-Zener body, with the additional inertial term. This model is proposed also to explain the experimental data (e.g., \citealt{Mats08a}; \citealt{LinEta10p}). Moreover, it is proved to be fundamental in thermodynamic rheology (see e.g., \citealt{Ver97b}; \citealt{Ful08a}; \citealt{FulEta14m}).

It is also remarkable, that the eliminated additional variable $r$ in eq. \re{frirheo}  has a clear and measurable physical meaning as recoverable displacement: it is responsible for the elastic properties of the body, representing deformations that relax to equilibrium, when the external force is zero. In this respect the recoverable displacement is an internal variable, spanning the non-equilibrium part of the state space, while the displacement $x$ spanning the equilibrium part.

\section{Results and discussion}

We will show to what extent the properties of rate- and state-dependent friction law can be reproduced by using our model. In this paper, we do not reproduce oscillations, therefore we neglect the inertial term with the coefficient $m$ in \re{lins1}. Assuming velocity control, the slip velocity $V$ is equal to the load-point velocity $V_l$. We will analyse in three steps; 1) with linear thermodynamic coefficients, 2) introducing nonlinearity by using a constant that is independent of load-point velocity, and 3) introducing nonlinear coefficient depending on load-point velocity. The features are as follows; 

(a)      instantaneous jump when the load-point velocity is abruptly changed,

(b)      the following relaxation to the stable sliding friction after the jump,

(c)      velocity strengthening,

(d)      velocity weakening,

(e)      nonlinear dependency of (a)-(d),

(f)      logarithmic dependency of (a)-(d),

(g)      offset of frictional coefficient independent from load-point velocity (e.g, $\mu=0.6-0.9$).

That is, (a)-(d) are qualitative features and (e)-(g) are quantitative ones.

\subsection{Linear thermodynamic coefficient}

In this subsection, we will show only equations and explain the relation between force and load-point velocity. The equations \re{lin1}-\re{lin2} reproduce that (a) instantaneous change of load-point velocity causes frictional jump and (b) following relaxation to the constant value of steady state friction. The stable-sliding frictional forces $F_{d1}$ and $F_{d2}$ at the load-point velocity $V_1$ and $V_2$ are 
\eqn{fd1}{
 F_{d1} &= \left( L_{1} -L_{12}\frac{L_{21}}{L_{2}}\right)V_{1}, \nl{fd2}
 F_{d2} &= \left( L_{1} -L_{12}\frac{L_{21}}{L_{2}}\right)V_{2},
}
with the following steady-state condition of eq. \re{lin2}
\eqn{steadyfr}{
   \dot r=L_{21}\dot x -L_2 kr=0.
}

The peak value at the instantaneous change from $V_1$ to $V_2$ is
\eqn{peakF}{
F_{dp} =F_{d1}+L_1\left( V_{2} -V_{1}\right)=L_1 V_2 -L_{12}\frac{L_{21}}{L_2}V_1.  
}

Therefore feature (c) can be also reproduced by using eqs. \re{fd1} and \re{fd2}, while properties (d)-(g) are not reflected in the model. E.g. the positive coefficient of the velocity difference between in eq. \re{fd1} and eq. \re{fd2} is required by the entropy inequality, therefore velocity weakening violates the second law in this framework. Accordingly, linear constant coefficients cannot reproduce all experimental data. 

{ The validity of our model in principle is not restricted to friction in case 
of slow speed jumps. It can be made compatible with the classical Dieterich and 
Ruina laws \re{Die_flaw}-\re{Rui_elaw} when introducing a transition of the 
force to the reference  velocity}.  If a constant value $F_{d0}$ is subtracted 
from $F_d$ in the left hand side of eqs. \re{fd1}-\re{fd2}, the equation of the 
steady-state friction for load-point velocity $V_{ss}$ is revised as
\eqn{fjump}{
F_{dss}=F_{d0}+\left( L_{1} -L_{12}\frac{L_{21}}{L_{2}}\right)V_{ss},
}
and it can reproduce feature (g). However, properties (d)-(f) cannot be reproduced. In order to improve our model we introduce nonlinearity in eqs. \re{fd1}-\re{fd2}.

\subsection{Velocity dependent nonlinear thermodynamic coefficient}

Here we will introduce a nonlinear thermodynamic coefficient in order to reproduce both the offset of frictional coefficient (feature (g)) and relative one depending on the load-point velocity (features (d)-(f)). The following nonlinearity is assumed for $L_1$, the damping coefficient:
\eqn{fe}{
L_{1}(V) =\frac{\beta}{1+\frac{\beta}{\alpha}\left\vert V \right\vert}.
}
The corresponding damping force $L_1(V) V$ is constant for $\frac{\alpha}{\beta} \ll \left\vert V \right\vert$  and  proportional to the velocity for $\frac{\alpha}{\beta} \gg \left\vert V \right\vert$. The term is switching between the two characteristic damping behaviors, the transition point is determined by the parameters $\alpha$ and $\beta$. This kind of nonlinearity was introduced in \cite{Van10a1} for modelling yield criteria and flow rules in thermodynamic plasticity, developing the ideas in \cite{HouPuz06b}.

In the following, we will show how eqs. \re{lin1}-\re{lin2} and \re{fe} can reproduce the experimental results. To compare the calculated result to experimental ones, the damping force, $F_d$, is related to frictional coefficient $\mu$,      
\eqn{dw}{
F_d =\mu N=\frac{\beta}{1+\frac{\beta}{\alpha}\left\vert V \right\vert} V - L_{12}kr,
}
where $N$ is the normal loading force. The variables and the thermodynamic coefficients are divided by $N$
\eqn{df}{
\mu=\frac{F_d }{N} =\frac{\beta/N}{1+\frac{\beta/ N}{\alpha/N}\left\vert V \right\vert} V - \frac{L_{12}}{N}kr.
}
and replaced as follows

\eqn{drel}{
\mu =\frac{\beta_{c}}{1+\frac{\beta_{c}}{\alpha_{c}}\left\vert V \right\vert} V - l_{12}kr
}
with $\alpha_c ={\alpha}/N,\,\ \beta_c ={\beta}/N,\,\ l_{12} =L_{12}/N$.

Finally we solve \re{drel} and \re{lin2}.
\begin{figure}
\begin{center}
\includegraphics[scale=.4]{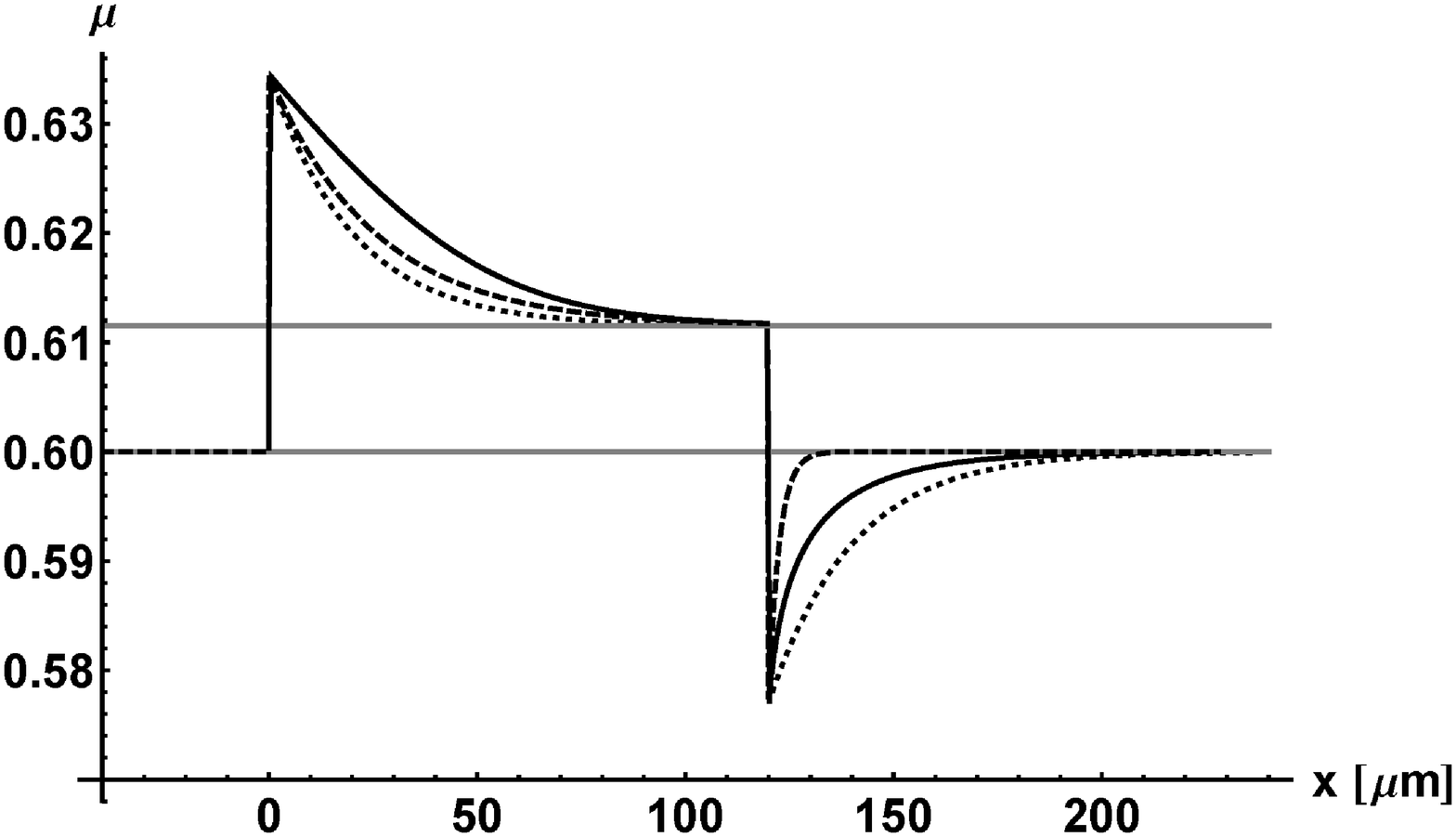}
\end{center}
\caption{ \label{fig_velstre}
Comparison of the rate- and state-dependent friction laws for velocity jumps 
and steady-state velocity strengthening experiments. The Dieterich model (solid 
line), Ruina model (dotted line) and the thermodynamic one (dashed line) show 
different relaxations. Dieterich and Ruina models are calculated with the 
parameters $V_* = 1\mu m/s$, $\mu_*=0.6$, $V_1 = 1\mu m/s$, $V_2 = 10\mu m/s$, 
$a=0.015$, $D_c=20\mu m$ and $b=0.01$. Thermodynamic model has the following 
parameters; $V_1 = 1\mu m/s$, $V_2 = 10\mu m/s$, $l_{12}=0.2/N$, 
$l_{21}=0.005$, $l_2=0.39 \mu m/sN$, $\alpha_c=0.64Ns/\mu m$, $\beta_c=10.0 
Ns/\mu m$, $k=1N/\mu m$.
}
\end{figure}
\begin{figure}
\begin{center}
\includegraphics[scale=.3]{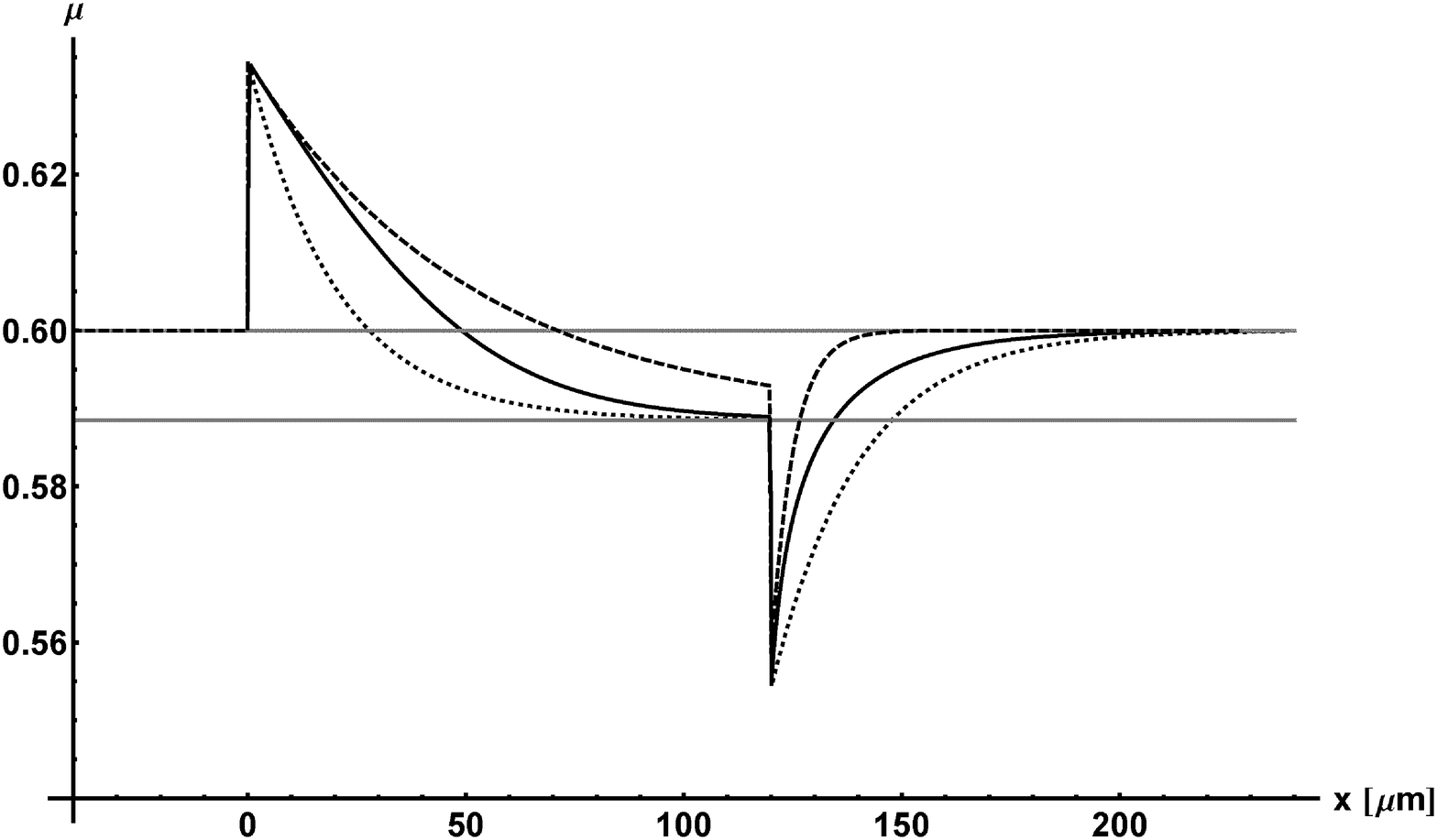}
\end{center}
\caption{ \label{fig_velweak}
Comparison of the rate- and state-dependent friction laws for velocity jumps 
and steady-state velocity weakening experiments. The Dieterich model (solid 
line), Ruina model (dotted line) and the thermodynamic one (dashed line) show 
different relaxations. Dieterich and Ruina models are calculated with the 
parameters $V_* = 1\mu m/s$, $\mu_*=0.6$, $V_1 = 1\mu m/s$, $V_2 = 10\mu m/s$, 
$a=0.015$, $D_c=20\mu m$ and $b=0.02$. Thermodynamic model has the following 
parameters; $V_1 = 1\mu m/s$, $V_2 = 10\mu m/s$, $l_{12}=0.2/N$, 
$l_{21}=0.005$, $l_2=0.195 \mu m/sN$, $\alpha_c=0.644Ns/\mu m$, $\beta_c=10.09 
Ns/\mu m$, $k=1N/\mu m$.
}
\end{figure}
In case of steady-state conditions, when the load-point velocity $V_{ss}$ is constant, the frictional coefficient is calculated substituting eq. \re{steadyfr} into \re{drel},
\eqn{RH5}{
\mu_{ss} =\left( \frac{\beta_{c}}{1+\frac{\beta_{c}}{\alpha_{c}}\left\vert V_{ss} \right\vert}- l_{12} \frac{l_{21}}{l_{2}}\right )V_{ss}, 
}
where $l_{21} =L_{21}$ and $l_{2} =L_{2}$.

We can derive the velocity dependency of steady-state friction by using eq. 
\re{RH5}.

The condition of velocity strengthening of steady-state friction is
\eqn{ineq1}{
  \frac{\beta_{c}}{\left(1+\frac{\beta_{c}}{\alpha_{c}}
   V_{1} \right)\left(1+\frac{\beta_{c}}{\alpha_{c}} V_{2} \right)} > 
   l_{12}\frac{l_{21}}{l_{2}}.
}

 On Fig. \ref{fig_velstre} we compare the performance of the classical Dieterich (eqs. \re{Die_flaw}-\re{Die_elaw} and Ruina (eqs. \re{Die_flaw} and \re{Nag_elaw}) models with our thermodynamic model (\re{drel} and \re{lin2}). We can see the frictional features ((a)-(c), (e) and (g)) are reproduced with the equations \re{drel} and \re{lin2}, while only feature (f) cannot be reproduced. The discrepancies between our model and the empirical laws can be seen in the type of relaxation. 

The condition of velocity weakening of steady-state friction is
\eqn{ineq2}{
  \frac{\beta_{c}}{\left(1+\frac{\beta_{c}}{\alpha_{c}}
   V_{1} \right)\left(1+\frac{\beta_{c}}{\alpha_{c}} V_{2} \right)} <
   l_{12}\frac{l_{21}}{l_{2}}.
}

 The calculation results are shown in Fig. \ref{fig_velweak}. The parameters of the classical models are from Marone (1998).
 Some parameters and initial values of the discrete thermodynamical model (\re{drel} and \re{lin2}) can be calculated from the steady-state values of friction: $l_2$, $\alpha_c$, and $\beta_c$. Other parameters are related to dynamic properties and are determined to show a comparable figure. 

We can see the simulated temporal change of friction with displacement showing 
velocity weakening, while the deviation remain in relaxation and f) logarithmic 
dependency of stable sliding friction to load-point velocity. These properties 
in velocity weakening are similar to that in velocity strengthening, and both 
of them are rooted in the constitutive equations (\re{drel} and \re{lin2})). 
Logarithmic dependency of load-point velocity requires a development of the 
model.

Both velocity strengthening and weakening are compatible with the thermodynamic 
conditions \re{tineq}, that is, they both satisfy  the second law of 
thermodynamics. Furthermore, the differences in requirements between eq. 
\re{ineq1} and eq. \re{ineq2} indicate a possible mechanism in this respect. In 
this model, it is caused by the difference between the linearity of the 
instantaneous jump and that of the following relaxation to the load-point 
velocity. A possible explanation of the mechanism can be the appearance of 
dissipative structures during frictional jump and the following relaxation. { 
There are several observations indicating substructures in sheared granular 
layers (see e.g. \cite{MaiMar99a}, figure 11). An indirect experimental 
observation of such phenomena may appear in the normal stress measurements (see 
e.g. \cite{LinDie92a}).} 

The difference in linear and non-linear relation of thermodynamic coefficients 
implies the difference in the microscopic structure in frictional layer that is 
formed of rock fragments between rock samples. It is inferred by the analogy of 
the differences in heat conduction and heat convection and the relation to the 
thermodynamic coefficients. In rock friction, loss of elastic stability and the 
development of shear bands can be fundamental. Thus the corresponding continuum 
description should introduce the loss of stability, a kind of plasticity, in 
case of an essentially granular media. Mesoscopic models { with weakly nonlocal 
constitutive relations} can provide an explanation, and also the exact form of 
the nonlinear thermodynamic coefficients.  It is an interesting problem and we 
will deal with it in a future work.

\section{Summary and conclusion}

Our simple thermodynamic model incorporates several aspects of rate- and state-dependent friction in a uniform theoretical framework. The separation of dissipative and non-dissipative parts, and the interpretation of the relaxation of state variable are remarkable. The most important advantage of our model is that the jump condition was a consequence of the evolution equation and was not a separate assumption. However, the number of the parameters is higher, and the logarithmic relaxation is not incorporated in our model. Introducing logarithmic terms in the theoretical framework as additional assumptions is straightforward. However, the most important question is to find a mechanism, an explanation of the origin of these terms. We suggest that a more detailed analysis of the continuum aspects may lead to such an explanation.

\section{Acknowledgement}   
 The authors thank to Tam\'as F\"ul\"op for valuable discussions. The work was supported by the grant Otka K81161 and K104260. N. Mitsui was supported by Canon Foundation in Europe.

\bibliographystyle{agsm}

\end{document}